ORCID: 0000-0002-2594-5559
**A. M. Gusak**
Bohdan Khmelnytsky National University of Cherkasy, 81 Shevchenko Blvd., UA-18031 Cherkasy, Ukraine
Ensemble3 Centre of Excellence, Wolczynska Str. 133, 01-919 Warsaw, Poland
amgusak@ukr.net

ORCID: 0009-0008-5158-1761
**Serhii Abakumov**
Bohdan Khmelnytsky National University of Cherkasy, 81 Shevchenko Blvd., UA-18031 Cherkasy, Ukraine
abakumov.serhii.official@gmail.com

# MODELING OF PATTERN FORMATION OF THE ORDERED INTERMEDIATE PHASES DURING CO-DEPOSITION OF BINARY THIN FILM

*Formation of the intermediate phase patterns in the thin-film co-deposition process is simulated using the Stochastic Kinetic Mean-Field method and Monte Carlo. Three basic morphologies of the 2D sections are distinguished: (1) spots (rod-like in 3D), (2) layered structures-lamellae, zigzags, and labyrinths (plate-like in 3D), and (3) net-like structures (inverse to spot-like structures, when spots become majority and the surrounding matrix becomes a minority). They are characterized and distinguished with the help of only one special topological parameter.*

## 1. Introduction.

Self-organization of two-phase structures (patterning) is a promising way of designing new materials for photonics, energy conversion, and accumulation [1–4]. Thus far, it has been mainly studied in the processes of directional eutectic crystallization, cellular precipitation, and spinodal decomposition [5–8]. In this paper, we examine the less-known case of self-organization: crystallization with pattern formation during co-deposition with a simultaneous reaction of two species from the vapor phase (for example, by sputtering or molecular beam epitaxy (MBE)). In particular, the co-deposition of immiscible components may provide binary materials with concentration modulations, that exhibit excellent mechanical properties [9–12]. Especially interesting for us is a paper [13] in which co-deposition by MBE led to decomposition with retained coherent boundaries between the emerging new phases. Another interesting example is related to ferromagnetic semiconductors and, in general, semiconductors doped with transition metals. For example, $Ge_{1-x}Mn_x$ films obtained during deposition (by MBE [14, 15] or magnetron sputtering [16]) demonstrate decomposition with the growth of self-organized nanocolumns or nanoprecipitates. In our study, we limited our simulations to the above-mentioned case when all boundaries between emerging phases remained coherent during co-deposition.

From a chemical point of view, the formation of new crystalline phases by decomposition and ordering represents a solid-state reaction during deposition. From a physical point of view, this is a sequence of phase transformations in an open system (the surface layer moves with the deposition rate) driven by the in-flux from the vapor phase and the out-flux into the crystalline phase. In our study, we limited ourselves to structural phase transformations during co-deposition, forming a rigid FCC lattice (changing only occupation probabilities at the fixed sites without changing the number and positions of sites). In principle, there are 3 possibilities: (1) decomposition into two solid solutions; (2) decomposition into a solid solution plus the ordered compound L12; and (3) decomposition into two different ordered compounds L12 and L10 (see Fig. 1(b,d)). Possibility (1) may be realized for alloys with positive mixing energy, corresponding decomposition cupola at the phase diagram, and corresponding W-shaped composition dependence of Gibbs free energy below the critical temperature. Possibilities (2,3) correspond to a phase diagram with three ordered intermediate phases (compounds), which are formed and have distinct boundaries with other phases due to a combination of negative mixing energies at the second coordination shell and positive mixing energy at the second coordination shell.

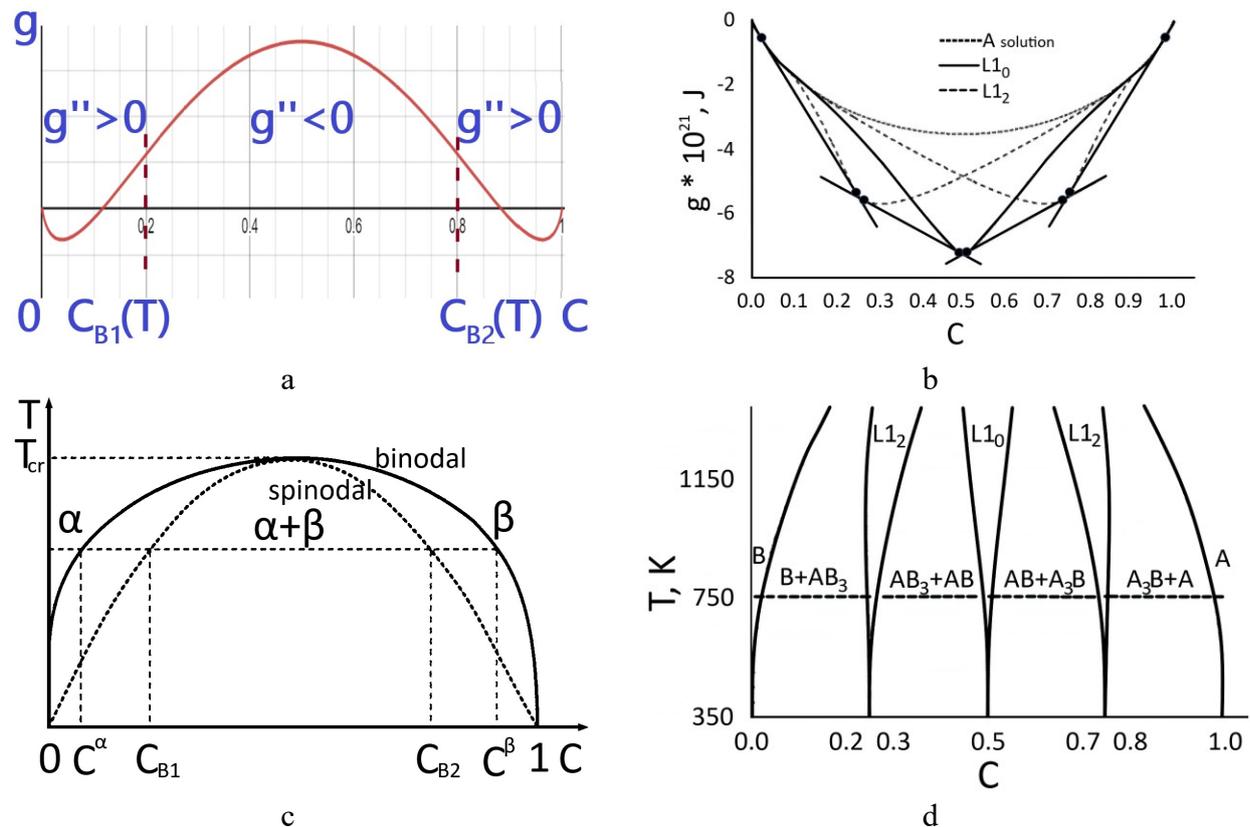

Fig. 1. Two possible types of binary systems used for co-deposition modeling in this paper are characterized by (a, b) composition dependencies of Gibbs free energy and (c, d) phase diagrams.

(a, c): positive mixing energy and corresponding W-shaped g(C)-curve leading to decomposition into two solid solutions (at least in the bulk, if the bulk diffusion is not frozen).

(b, d): negative mixing energy for the nearest neighbors and positive mixing energy for the next nearest neighbors, leading to the ordering of compounds within narrow concentration ranges around 1/4, 1/2, and 3/4, and leading to decomposition into A3B compound +A(B) solution or A3B compound + AB compound, etc., beyond the mentioned narrow ranges.

In the recent Letter [17] the corresponding author of this paper jointly with A.Titova obtained some first results in case 1 (positive mixing energy) concerning, first of all, the qualitative behavior of patterns with variation of composition and initial (preexisting) patterns. We found that under the condition of completely frozen bulk diffusion, the 2D morphology in the atomic planes normal to the deposition/growth direction can be classified using two basic patterns: spot-like (rods in 3D pictures) and layered (lamellar, zigzag, or labyrinth), as well as mixed morphology (spots between layers or some layers between spots). Such mixed morphology was obtained (in our simulations) for the composition intervals (0.39-0.44) and (0.66-0.71), for all deposition rates. The influence of the initial conditions on the resulting steady-state pattern was found only within these co-deposition composition intervals.

Decomposition in general and during co-deposition may proceed in two different cases
1. Species A and B "don't like each other", which in the case of the regular solid solution means positive mixing energy at the first coordination shell and existence of spinodal decomposition region (instability in respect to smallest composition fluctuations) - see Fig.1a,c. In this case, the decomposition is a direct consequence of energetic "dislike".
2. Species A and B "like each other" (negative mixing energy within the first coordination shell), especially in some special (stoichiometric) proportions, leading to division into sublattices, providing the maximum possible number of AB interactions (bonds) within the first coordination shell. The tendency to order at some specific stoichiometric compositions becomes especially distinct if the mixing energy in the second coordination shell has the opposite sign (positive). In this case, 2 the decomposition is a consequence of the tendency to optimize (maximize) the number of AB bonds by the precipitation of one or both ordered phases with different compositions (see Fig. 1(b,d)).

As mentioned above, we will work with a rigid FCC lattice. In binary cases, the following three ordered phases are possible: A3B, AB3 with structure $L1_2$ (for example, an ideally ordered A3B phase corresponds to A occupying the face centers and B occupying the vertexes of elementary cubic cells), AB with structure $L1_0$ (for example, atomic layers (001) periodically occupied by A and B).

For modeling case 2, we need at least a negative mixing energy in the first coordination shell. However (as mentioned above), to distinguish between phases A3B, AB, and AB3 (larger concentration gaps between them at the phase boundary), it is convenient to add positive mixing energy to the interactions with the second coordination shell. We realize this and check it below.

We model the situation when the bulk diffusion is frozen (temperature lower than $0.3T_{melt}$ in the case of a single-component system), so that the deposited atoms can move only within the surface (before being buried by the deposition flux), or maximum by several atomic planes inside. In this paper, we focus mainly on the morphology of the resulting alloys after co-deposition and reaction - in particular, on the dependence of this morphology on composition $C^{dep}$ of the incoming deposition flux and on the ratio of deposition and surface diffusion rates. Such dependence will be studied using a modification of the atomistic-scale stochastic kinetic mean-field (SKMF) method (modification is described in Section II) and, alternatively, by the standard Monte Carlo (MC) method. After this, we start the numerical modeling of the topology evolution. One of the ways to quantify the morphology is to introduce the special topological parameter of the 2D sections of the deposited thin film (Section III). The simulation of pattern formation in case 2 (negative mixing energy at the first coordination shell and positive mixing energy at the second coordination shell) is presented in Section IV by KMF and by MC in Section V.

## 2. Modification of Stochastic Kinetic Mean-Field approximation (SKMF) for open systems - the surface layer during co-deposition.

In numerical simulations we will keep in mind the following simplified picture of layer-by-layer co-deposition: Let first the new atomic layer be deposited completely instantaneously, and only after this, we "switch on" the surface diffusion for some fixed number M of time steps. The diffusion proceeds via an artificial (but kinetically effective in simulations) exchange mechanism within a non-linear "almost 2D" version of the recent development of the stochastic kinetic mean-field model SKMF [18-23]. (This model was, in turn, the development of a quasi-one-dimensional KMF model [24-26] including the dynamic noise of the micro fluxes between the neighboring sites). The basic algorithm and free software can be found in skmf.eu.

Interatomic exchanges are only permitted within this new plane, but the interaction energies are considered for the nearest neighbors within this plane and just below this plane. The number M of time steps for surface diffusion before the deposition of the new atomic layer, at a fixed time step, is inversely proportional to the deposition rate. Mathematically, as in all kinetic mean-field models, we numerically solved the set of master equations of occupancy probabilities for all atomic sites:

The master equation for occupancy probabilities in the sites of a rigid lattice and with an account of influx and out-flux is self-consistent and non-linear (frequencies exponentially depend on energies, and energies are the linear functions of probabilities):

$$\frac{\partial C_A}{\partial t} = \sum_{in=1}^{Z^{\parallel}} \{ -C_A(i)C_B(in)\Gamma_{AB}(A(i) \leftrightarrow B(in)) + C_B(i)C_A(in)\Gamma_{AB}(A(in) \leftrightarrow B(i)) \} \quad (1)$$

$$0 < t < Mdt = \frac{\delta}{V}$$

The master equation (1) for the occupation probabilities is written only for the surface layer, $Z^{\parallel}$ being the number of nearest sites for exchange within the surface atomic layer. Summation in eq. (1) is made only over sites within the same surface (top) atomic layer. In the case of the (001)-plane orientation of the FCC lattice one has ($1 \leq in \leq Z^{\parallel} = 4$). $\delta$ is the thickness of the top atomic layer, within which only the exchanges are possible. For the (001)-plane orientation, $\delta = \frac{a}{2}$

Frequencies depend on the energetic barrier, which is the difference between the saddle-point energy (which is not fixed explicitly but typically assumed to be common for all jumps), and the known energy before the jump (atomic exchange) ($E_A(i) + E_B(in)$)

$$\Gamma_{AB}(A(i) \leftrightarrow B(in)) = v_0 exp(-\frac{E^s - (E_A(i) + E_B(in))}{kT}) \quad (2)$$

Energies are calculated in mean-filed approximation:

$$E_A(i) = \sum_{i'=1}^{Z^{\parallel}+Z^{\perp}} (C_A(i')V_{AA} + C_B(i')V_{AB}) = (Z^{\parallel} + Z^{\perp})V_{AB} + (V_{AA} - V_{AB})\sum_{i'=1}^{Z^{\parallel}+Z^{\perp}} C_A(i') \quad (3)$$

$$E_B(in) = \sum_{in'=1}^{Z^{\parallel}+Z^{\perp}} (C_A(in')V_{BA} + C_B(in')V_{BB}) = (Z^{\parallel} + Z^{\perp})V_{BB} + (V_{AB} - V_{AA})\sum_{in'=1}^{Z^{\parallel}+Z^{\perp}} C_A(in')$$

$$(4)$$

In the case of co-deposition of the (001) planes of the FCC lattice, as mentioned above, the number of nearest neighbors within the top plane (simultaneously, the number of possible atomic exchanges) is $Z^{\parallel} = 4$, and the number of nearest interacting neighbors in the preceding (subsurface) plane is $Z^{\perp} = 4$. "i" and "in" are two neighboring sites withing the top atomic plane exchanging by atoms. At fixed "i" there are $Z^{\parallel} = 4$ possibilities for "in". "i'" are the nearest interacting neighbors of the site "i", and their number is $Z = Z^{\parallel} + Z^{\perp} = 8$, in' are the nearest interacting neighbors of the site in, and their number is also $Z = Z^{\parallel} + Z^{\perp} = 8$.

Below we take for simplicity: $V_{AA} = 0, V_{BB} = 0, V_{AB} = E_{mix}$. Then

$$\Gamma_{AB}(A(i) \leftrightarrow B(in)) = v_0 e^{-E^s/kT} exp[\frac{E_{mix}}{kT}(Z - \sum_{i'=1}^{Z} C_A(i') + \sum_{in'=1}^{Z} C_A(in'))] \quad (5)$$

$$\Gamma_{AB}(A(in) \leftrightarrow B(i)) = v_0 e^{-E^s/kT} exp[\frac{E_{mix}}{kT}(Z - \sum_{in'=1}^{Z} C_A(in') + \sum_{i'=1}^{Z} C_A(i'))] \quad (6)$$

Then

$$\frac{\partial C_A(i)}{\partial tt} = \sum_{in=1}^{Z^{\parallel}} \{ -C_A(i)(1 - C_A(in)) \cdot exp[\frac{E_{mix}}{kT}(\sum_{in'=1}^{Z} C_A(in') - \sum_{i'=1}^{Z} C_A(i'))] + (1 - C_A(i))C_A(in)exp[\frac{E_{mix}}{kT}(\sum_{i'=1}^{Z} C_A(i') - \sum_{in'=1}^{Z} C_A(in'))] \} \quad (7)$$

where the non-dimensional time is

$$tt = t \cdot v_0 exp[\frac{ZE_{mix} - E^s}{kT}] \quad (8)$$

Modification of the SKMF for decomposition with ordering typically requires interactions within two coordination shells and atomic exchanges, as a minimum, between two atomic planes. This is analyzed in Section IV.

**3. Quantification of 2D patterns topology.**

During co-deposition, the morphology of the 2D sections at first changes with increasing film height but eventually tends to an almost fixed (steady-state) pattern. As mentioned above and as we checked recently for spinodal decomposition under co-deposition [17], the observed topologies can be classified as spot-like (rod-like in 3D) – Fig.2a, layered (lamellar, zigzag, or labyrinth-like) – Fig.2b1,b2,b3, and their mixture. The spot-like morphology corresponds to isolated precipitates of the minority phase surrounded by a common percolation cluster (matrix) of the majority phase (Fig. 3a). In the layered morphology the components are more or less equivalent to each other (except for volume fractions, which depend on the ratio of the layer thicknesses). We will see below (simulating the topology during co-deposition with simultaneous decomposition into two ordered phases A3B and AB) that one more topology is possible - we will call it "net-like" (like a net for fishing, see below) – Fig.3c. In this structure, the majority and minority change place in comparison with spot-like morphology: the majority phase is closed in cells, and the minority forms thin walls around these cells.

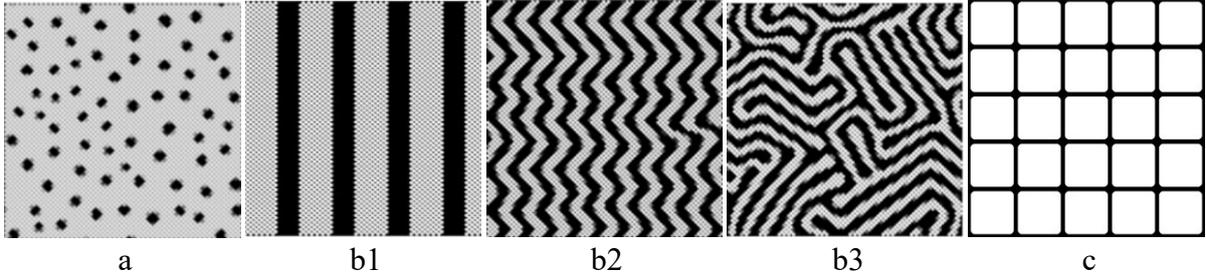

a         b1         b2         b3         c

*Fig. 2. Three "pure" morphologies for the 2D-sections: (a) isolated spots of minority phase surrounded by a common cluster of "majority" phase; (b) layered structure (lamellae or zigzag or labyrinth) without explicit preference for minority or majority phase; (c) net-like (cell) structure which is in some sense inverse to spot-like.*

To quantify the morphology of the 2D patterns, we introduced in recent paper [17] a new (topological) parameter P. This parameter is related to the size dependence of the number of clusters of some phase within a subregion of the entire sample in the subregion area. Let all subregions be the squares h·h with all possible positions inside the square sample L·L. Let N($h$) be the mean number of clusters of the "minority" phase (with a fraction not larger than half) in the sub-region, averaged over all possible positions of the subregion.
By our definition,

$$P = \frac{d\ln N}{d\ln h} \quad (9)$$

If $h$ is larger than the characteristic length of the phase separation, then in the case of regular (marginal, ideal) structures we have for the minority phase:
1. Ideal rod structure: $N = h^2/\lambda^2 \rightarrow \ln N = 2\ln h + \text{const} \rightarrow P_{minority} = d\ln N/d\ln h = 2$.
2. Ideal periodic lamellae: $N = h/\lambda \rightarrow \ln N = \ln h + \text{const} \rightarrow P_{minority} = d\ln N/d\ln h = 1$.
3. Ideal net structure for minority (or cell structure for majority): The ideal net in 2D is one large cluster. Therefore, for all h larger than the size of the net cells, one has $N = 1 = \text{const} \rightarrow \ln N = 0 = \text{const} \rightarrow P_{minority} = d\ln N/d\ln h = 0$.

We introduce the topological parameter not only for the minority phase but also for the majority phase.
It is easy to check that for majority phase one has:

1. The ideal rod structure of the minority means that the majority phase around the rods is one large cluster: N = 1, so that $P_{majority} = d\ln 1/d\ln h = 0$.
2. Ideal periodic lamellae: majority lamellae are thicker, but their number is the same as that of minority phase lamellae: $N = h/\lambda \rightarrow \ln N = \ln h + const \rightarrow P_{majority} = d\ln N/d\ln h = 1$.
3. Ideal net structure for minority (or cell structure for the majority): if cells are isolated from one another by the surrounding net intervals, the calculation of the cluster number is analogic to the number of rods. So, in this case, $N = h^2/\lambda^2 \rightarrow \ln N = 2\ln h + const \rightarrow P_{minority} = d\ln N/d\ln h = 2$.

It should be noted that at least in the above-mentioned three ideal cases, $P_{minority} + P_{majority} = 2$. Is this relation more universal, may we introduce a new "principle of complementarity" - that we do not know so far. Our first simulations for limited sizes confirm only the tendency but not the strict equation. The case of asymptotics for extremely large samples will be studied elsewhere.

**4. Patterns formation during co-deposition with precipitation of one or both ordered phases at various compositions 0 < C < 0.25 (one of the phases is ordered) and 0.25 < C < 0.50 (both phases are ordered) at various deposition-to-diffusion rates: Kinetic Mean-Field simulation.**

Here we will use the modification of KMF, suggested in [19] for ordering and diffusion in the ordered phases, and in [20] for the formation and competition of intermediate ordered phases during reactive diffusion. First of all, we should take into account the interactions at least within the second coordination shell, as was done in [20], to guarantee good distinction and broad concentration gaps between the narrow concentration ranges of the ordered compounds. Second, to provide the possibility of full ordering during deposition (including exchanges between sublattices), we broadened the kinetic possibilities for surface and subsurface atoms: in our model, they may exchange with neighbors not only within the top plane but also with (as a minimum) the atoms in the plane below the top plane. This means that instead of one type of equation (1), in this case, we write down and solve three different equations for the first (top) layer (exchanges with 4+4=8 neighboring sites within planes 1 and 2), for the second layer (exchanges with 4+4+4=12 sites within planes 1,2 and 3), and for the third layer (exchanges with 4+4=8 sites within planes 2 and 3). The energies are now calculated in the mean-filed approximation with an account of the second coordination shell. For example, in the case of both exchanging atoms in the top (first) layer.

$$E_A(i) = \sum_{i'_1=1}^{Z_1^{\parallel}+Z_1^{\perp}} \left(C_A(i'_1)V_1^{AA} + C_B(i'_1)V_1^{AB}\right) + \sum_{i'_2=1}^{Z_2^{\parallel}+Z_2^{\perp}} \left(C_A(i'_2)V_2^{AA} + C_B(i'_2)V_2^{AB}\right) =$$

$$= (Z_1^{\parallel} + Z_1^{\perp})V_1^{AB} + (V_1^{AA} - V_1^{AB})\sum_{i'_1=1}^{Z_1^{\parallel}+Z_1^{\perp}} C_A(i') + (Z_2^{\parallel} + Z_2^{\perp})V_2^{AB} + (V_2^{AA} - V_2^{AB})\sum_{i'_2=1}^{Z_2^{\parallel}+Z_2^{\perp}} C_A(i')$$

$$E_B(in) = \sum_{in'_1=1}^{Z_1^{\parallel}+Z_1^{\perp}} \left(C_A(in'_1)V_1^{BA} + C_B(in'_1)V_1^{BB}\right) + \sum_{in'_2=1}^{Z_2^{\parallel}+Z_2^{\perp}} \left(C_A(in'_2)V_2^{BA} + C_B(in'_2)V_2^{BB}\right) =$$

$$= (Z_1^{\parallel} + Z_1^{\perp})V_1^{BB} + (V_1^{AB} - V_1^{BB})\sum_{in'_1=1}^{Z_1^{\parallel}+Z_1^{\perp}} C_A(in'_1) + (Z_2^{\parallel} + Z_2^{\perp})V_2^{BB} +$$

$$+(V_2^{AB} - V_2^{BB}) \sum_{in'_2=1}^{Z_2^{\parallel}+Z_2^{\perp}} C_A(in'_2)$$

(10)

$i'_1, i'_2$ are the indexes of nearest neighbors and next-nearest neighbors of site "i" respectively, $in'_1, in'_2$ are the indexes of nearest neighbors and next-nearest neighbors of the site "in".

The nondimensional time is

$$tt = t \cdot \nu_0 exp\left[\frac{Z_1 E_1^{mix} + Z_2 E_2^{mix} - E^s}{kT}\right]$$

(11)

Simulation parameters: $\frac{E_1^{mix}}{kT} = -0.2802$ within the first coordination shell and 0.6531 for the second coordination shell. The simulation was performed according to eqs. (1-8) at various deposition-to-diffusion rates $M = \frac{\delta}{Vdt} = \frac{1}{\upsilon \cdot dtt}$ (M is the number of diffusion time steps per deposited layer).

### *4.1.* Main results, 0 < C < 0.25 (one phase is a solid solution of B in A, another phase is a compound A3B), KMF:

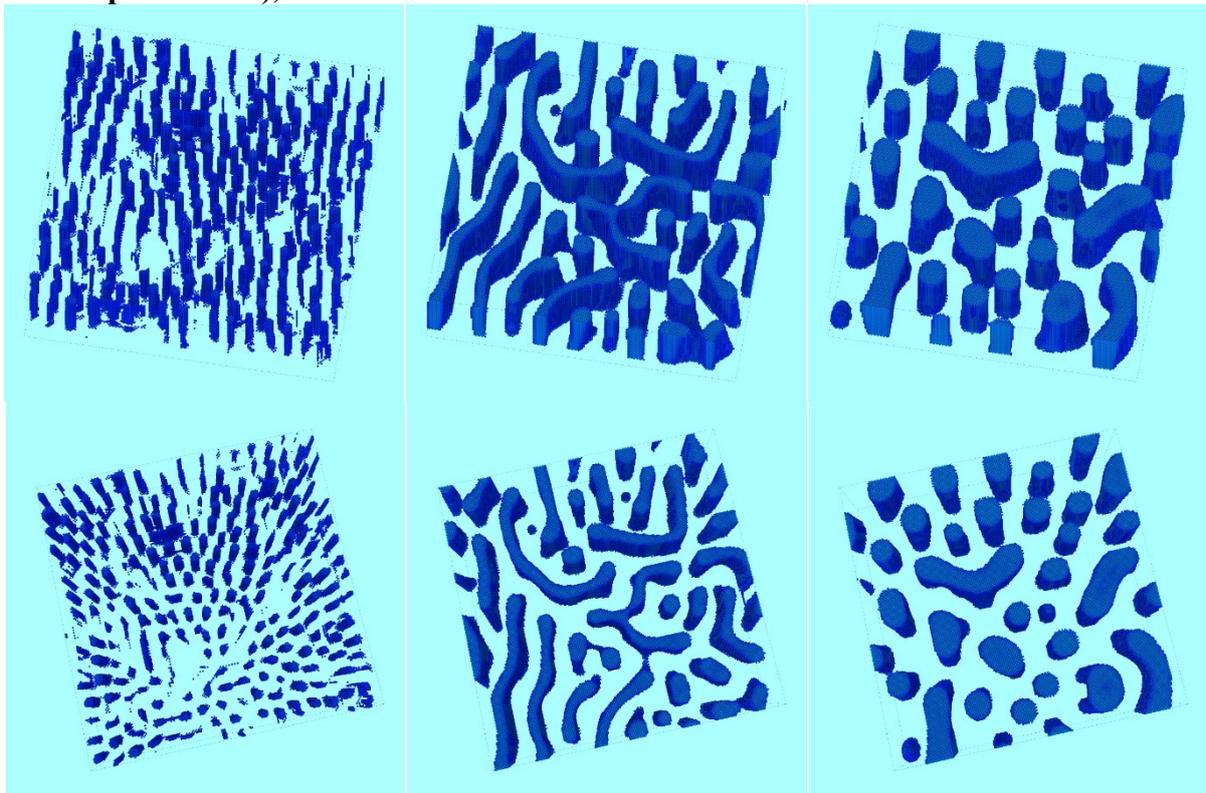

*Initial Cdep: 10.0 ± 1%; Exchange planes: 3; Medium-Nodal Cdep > 15%; (Minority phase)*
*Approximation: LnN(h) = P * ln(h) + const*

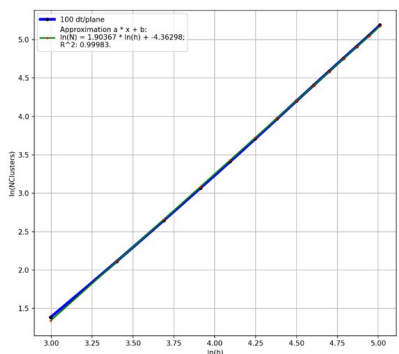 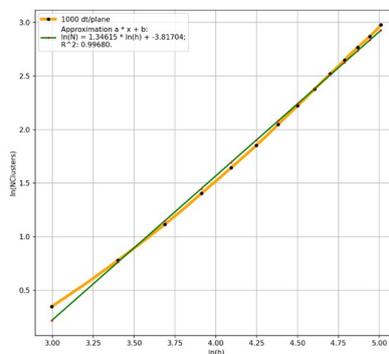 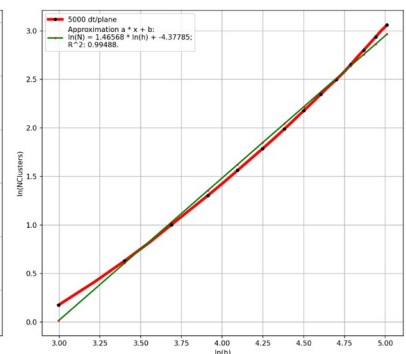

| | | |
|---|---|---|
| *LnN(h) = 1.90367 * ln(h) − 4.36298, P = 1.90; R^2: 0.99983* | *LnN(h) = 1.34615 * ln(h) − 3.81704, P = 1.35; R^2: 0.99680* | *LnN(h) = 1.46568 * ln(h) − 4.37785, P = 1.47; R^2: 0.99488* |

A

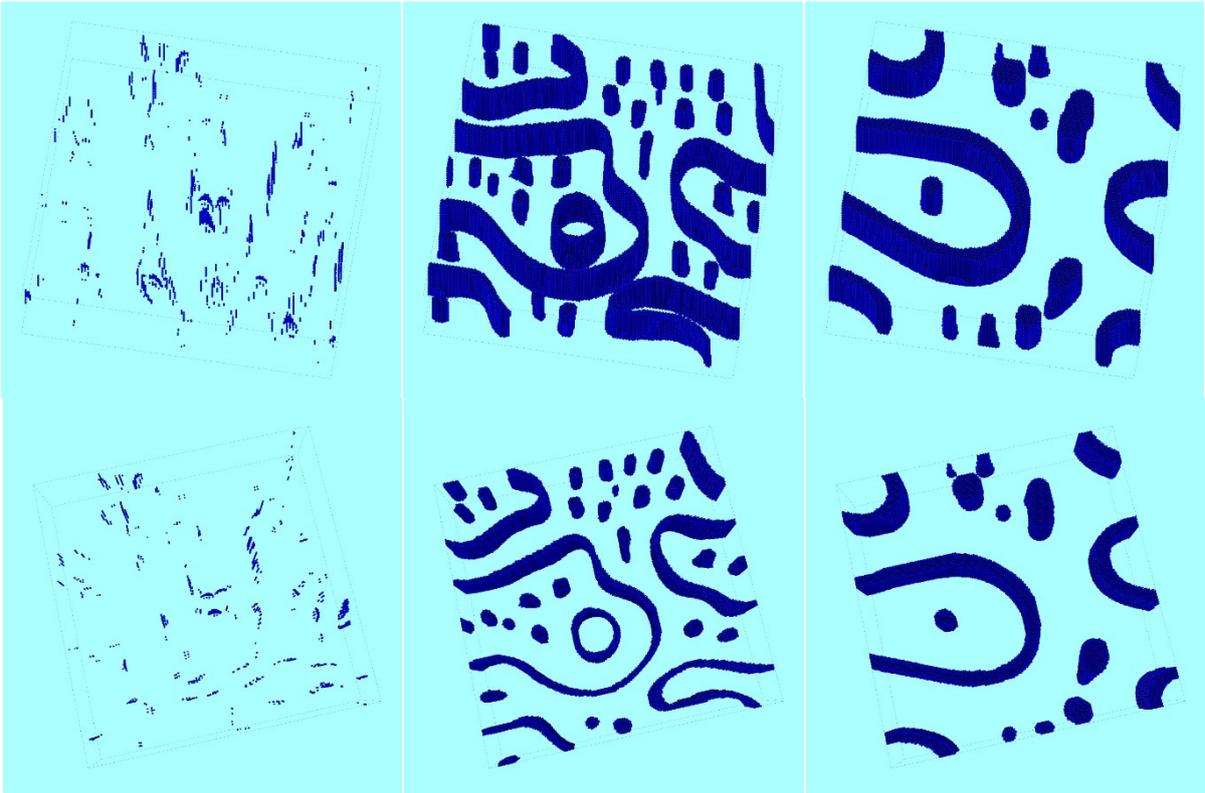

*Initial Cdep: 20.0 ± 1%; Exchange planes: 3; Medium-Nodal Cdep < 10%; (Minority phase)*
*Approximation: LnN(h) = P * ln(h) + const*

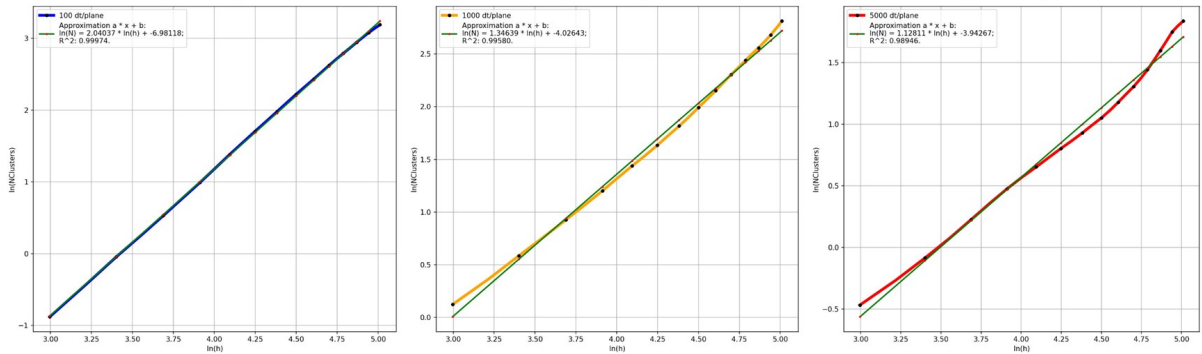

| LnN(h) = 2.04037 * ln(h) − 6.98118, P = 2.04; R^2: 0.99974 | LnN(h) = 1.34639 * ln(h) − 4.02643, P = 1.35; R^2: 0.99580 | LnN(h) = 1.12811 * ln(h) − 3.94267, P = 1.13 R^2: 0.98946 |

b

*Fig. 3. Patterns of A3B-phase within A-matrix (a) and patterns of A within A3B-matrix (b) and corresponding plots $\frac{\ln(N)}{\ln(h)}$ for calculation of the topological parameter P. Each pattern is shown in two different projections. Sample 200x200x55. Deposition rate/diffusion rate 100 dt/plane (left), 1000 dt/plane (center), 5000 dt/plane (right). Exchanges are permitted within 3 upper planes after deposition*

  a. $C^{dep}$ : $10.00 \pm 1\%$ ; sites with mean B-fraction $C^{mean} > 0.15$ are marked- it corresponds to phase A3B inclusions within the matrix of A-based solid solution.
  b. $C^{dep}$ : $20.00 \pm 1\%$ ; sites with mean B-fraction $C^{mean} < 0.10$ are marked- it corresponds to A-based solution inclusions within the matrix of A3B.

### 4.2. Main results, 0.25 < C < 0.50 (both phases A3B and AB are ordered compounds), KMF:

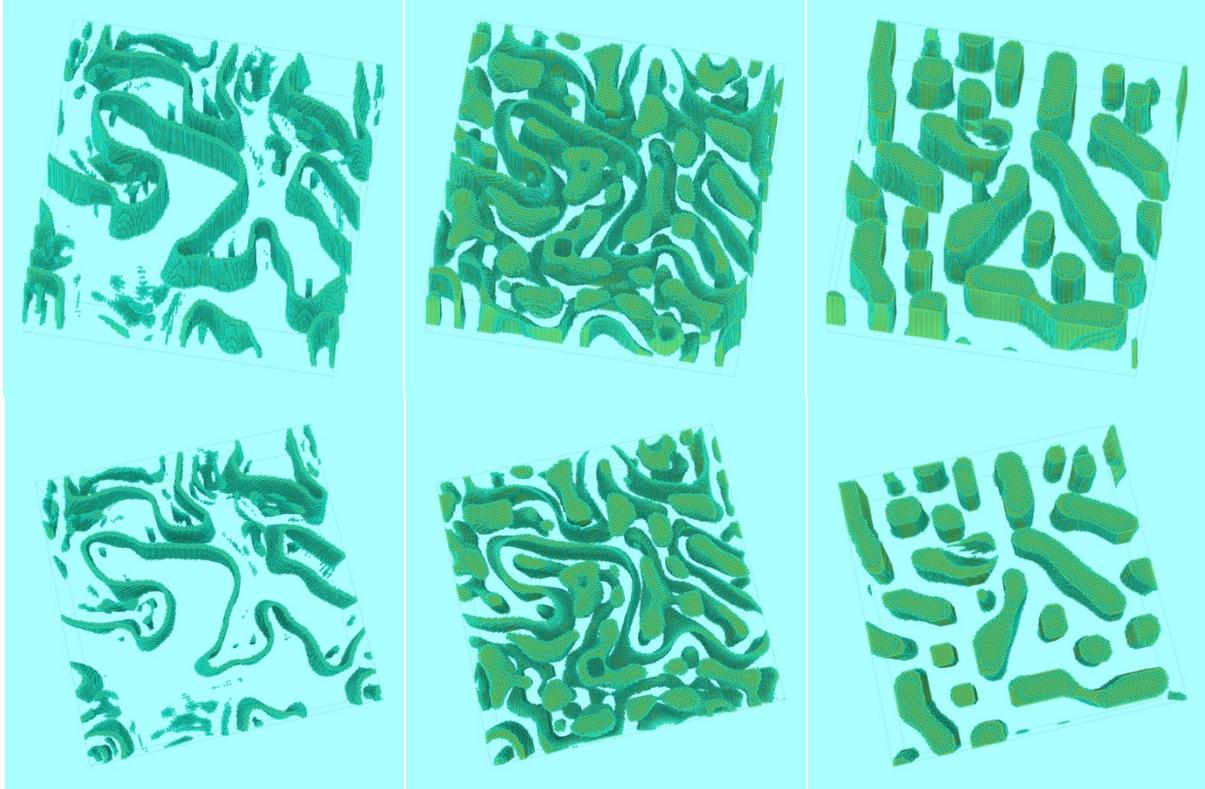

*Initial Cdep: 35.0 ± 1%; Exchange planes: 3; Medium-Nodal Cdep > 40%; (Minority phase)*
*Approximation: LnN(h) = P * ln(h) + const*

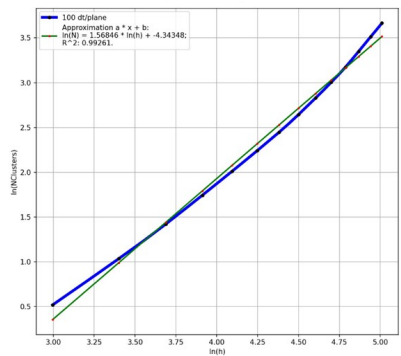 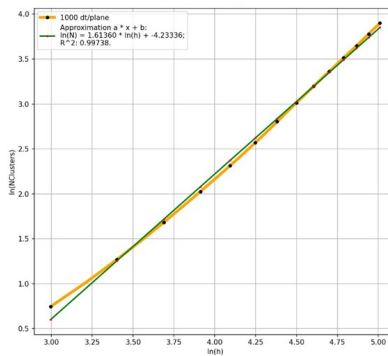 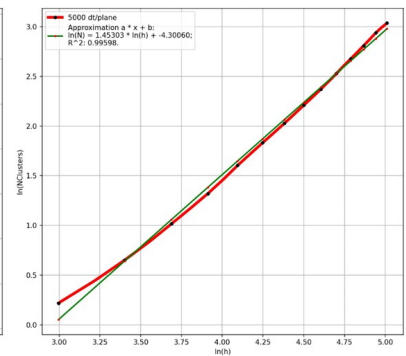

*LnN(h) = 1.56846 * ln(h) - 4.34348, P = 1.57; R^2: 0.99261*  *LnN(h) = 1.61360 * ln(h) - 4.23336, P = 1.61; R^2: 0.99738*  *LnN(h) = 1.45303 * ln(h) - 4.30060, P = 1.45; R^2: 0.99598*

a

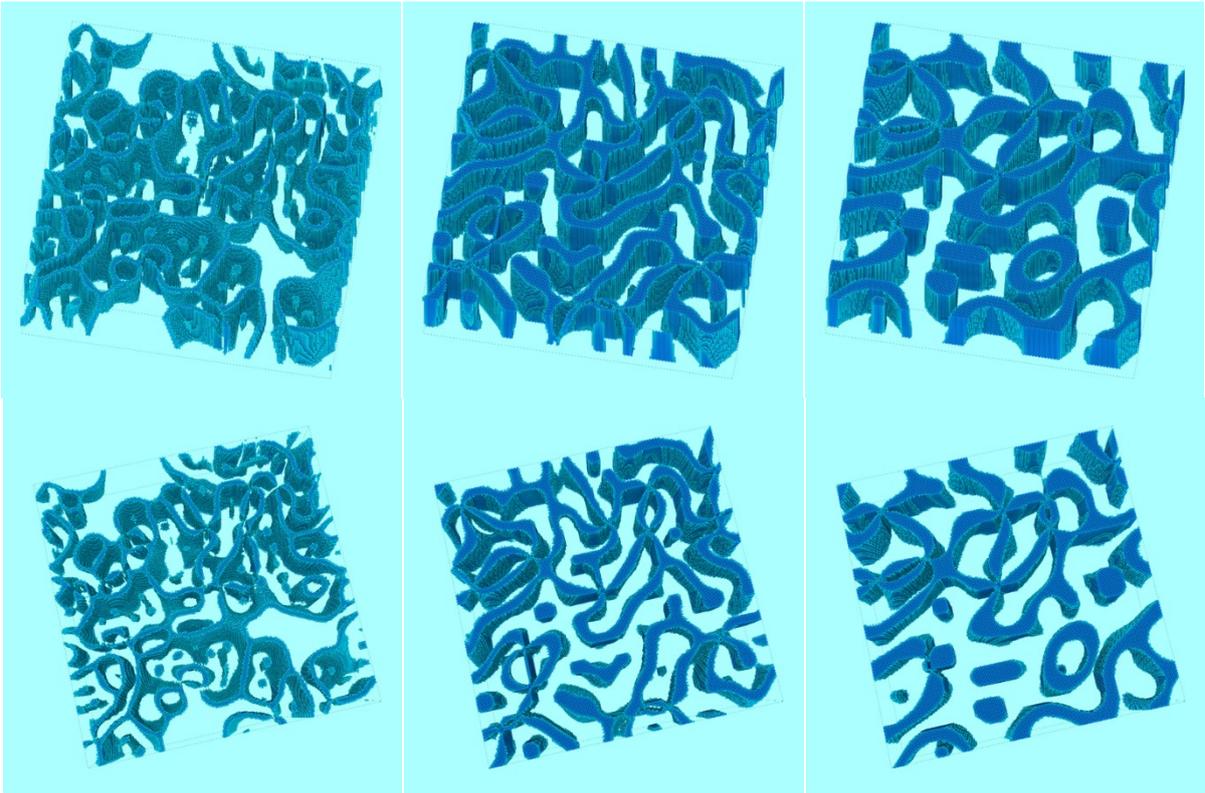

*Initial Cdep: 40.0 ± 1%; Exchange planes: 3; Medium-Nodal Cdep < 35%; (Minority phase)*
*Approximation: LnN(h) = P * ln(h) + const*

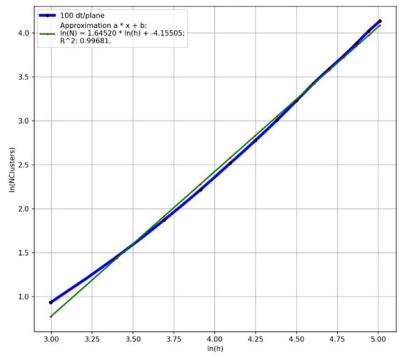
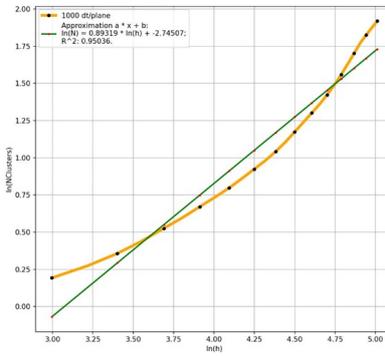
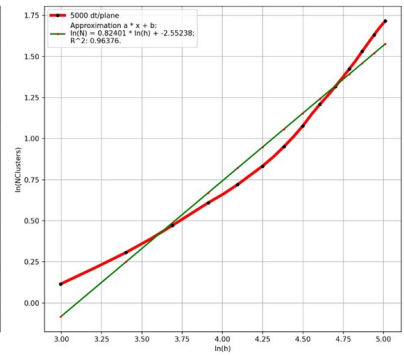

*LnN(h) = 1.64520 * ln(h) − 4.15505, P = 1.64; R^2: 0.99681*

*LnN(h) = 0.89319 * ln(h) − 2.74507, P = 0.89; R^2: 0.95036*

*LnN(h) = 0.82401 * ln(h) − 2.55238, P = 0.82; R^2: 0.96376*

b

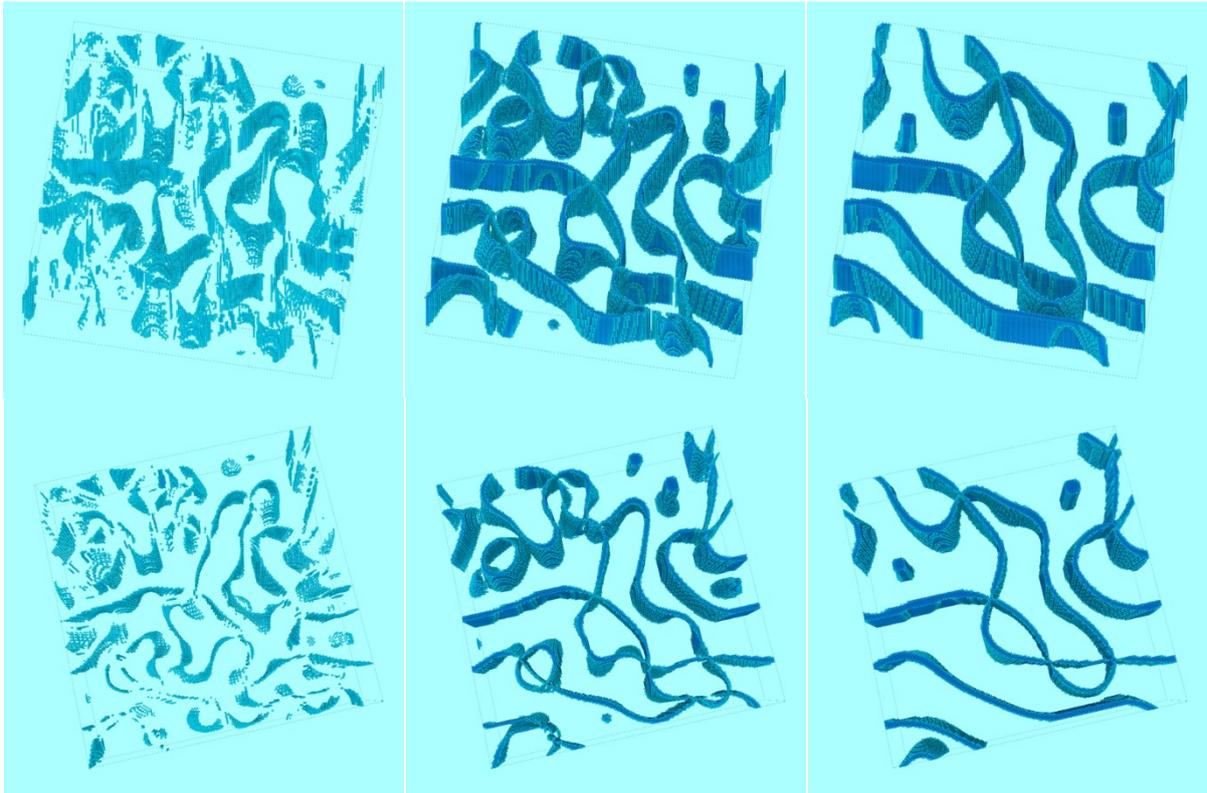

*Initial Cdep: 45.0 ± 1%; Exchange planes: 3; Medium-Nodal Cdep < 35%;*
*(Minority phase)*
*Approximation: LnN(h) = P * ln(h) + const*

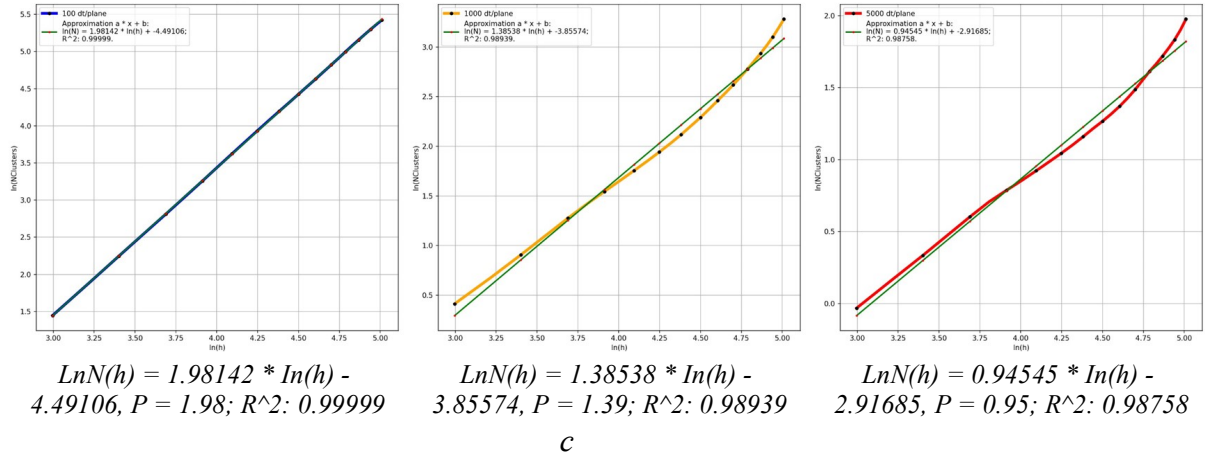

*LnN(h) = 1.98142 * ln(h) - 4.49106, P = 1.98; R^2: 0.99999*

*LnN(h) = 1.38538 * ln(h) - 3.85574, P = 1.39; R^2: 0.98939*

*LnN(h) = 0.94545 * ln(h) - 2.91685, P = 0.95; R^2: 0.98758*

*c*

*Fig. 4. Patterns of AB-phase within A3B-matrix (a) and patterns of A within A3B-matrix and corresponding plots $\frac{\ln(N)}{\ln(h)}$ for the calculation of the topological parameter P. Each pattern is shown in two different projections. Sample 200x200x55. Deposition rate/diffusion rate 100 dt/plane (left), 1000 dt/plane (center), 5000 dt/plane (right) Exchanges are permitted within three upper planes after deposition*

   a. $C^{dep}$ : $35.00 \pm 1\%$ ; sites with mean B-fraction $C^{mean} > 0.40$ are marked; it corresponds to minority phase AB within the majority matrix of A3B.
   b. $C^{dep}$ : $40.00 \pm 1\%$ ; sites with mean B-fraction $C^{mean} < 0.35$ are marked; it corresponds to minority A3B within the majority matrix of AB.
   c. $C^{dep}$ : $45.00 \pm 1\%$; sites with mean B-fraction $C^{mean} < 0.35$ are marked; it also corresponds to minority A3B within the majority matrix of AB.

Note that at compositions of 0.40 and 0.45, the parameter becomes lower than 1: it correlates with some elements of net-like morphology for minority phase A3B (cell-like for majority phase AB), which can be seen in Fig. 4b,c.

# 5. Pattern formation during co-deposition with precipitation of one or both ordered phases at various compositions 0 < C < 0.25 (one of the phases is ordered) and 0.25 < C < 0.50 (both phases are ordered) at various deposition-to-diffusion rates: Monte Carlo simulation.

Simulation parameters: $\frac{E_1^{mix}}{kT} = -0.2802$ within the first coordination shell, and 0.6531 for the second coordination shell. Where M is the number of Monte Carlo steps per deposited layer.

We used the Metropolis algorithm with exchange and interaction constraints that are the same as those for the kinetic mean-field method (described in the previous section).

In the Monte Carlo method, each site is occupied by A or B, so "microscopic composition" has only two choices for C(i): 0 or 1 (contrary to the KMF method, which has a continuous composition range for occupancy probabilities). This simple fact makes all kinds of averaging of the concentration and order parameter over surrounding clusters rather discreet.

We use $C^{mean}(i)$ -composition averaged over a cluster consisting of 1+12 sites:

$$C^{mean} = \frac{C(i) + \frac{1}{4}\sum_{in=1}^{12} C(in)}{4} \qquad (12)$$

Since $C(i) = \pm 1$, the $C^{mean}$ may be equal only to k/16, with integer k ranging from 0 to 16. One may easily check that this definition of $C^{mean}$ provides $C_A^{mean} = \frac{12}{16}$ for any site of the ordered stoichiometric compound A3B (structure L1$_2$), $C_A^{mean} = \frac{4}{16}$ for any site of the ordered stoichiometric compound AB3 (structure L1$_2$), and $C_A^{mean} = \frac{8}{16}$ for any site of the ordered stoichiometric compound AB (structure L1$_0$).

We also introduced (according to [16]) the "Local Long-Range Order parameter" for structures L1$_2$ and L1$_0$.

For L1$_2$,

$$\eta(i) = \max\{abs(\eta_{1,2,3}^I(i)), abs(\eta_4^{II}(i)), \qquad (13)$$

Where

$$\eta_{1,2,3}^I(i) = \frac{\frac{C(i) + \frac{1}{4}\sum_{in=1}^{8} C(in)}{3} - \frac{C(i) + \frac{1}{4}\sum_{in=1}^{12} C(in)}{4}}{1 - 3/4}, \quad \eta_4^{II}(i) = \frac{C(i) - \frac{C(i) + \frac{1}{4}\sum_{in=1}^{12} C(in)}{4}}{1 - 1/4} \qquad (14)$$

Cases 1,2, and 3 correspond to three different ways of choosing the eight nearest neighbors out of 12 (in two orthogonal planes out of three orthogonal planes (100), (010), and (001))

For L1$_0$,

$$\eta(i) = \max\{abs(\eta_{1,2,3}^I(i))\}, \qquad (15)$$

Where

$$\eta_{1,2,3}^I(i) = \frac{\frac{C(i) + \frac{1}{4}\sum_{in=1}^{4} C(in)}{2} - \frac{C(i) + \frac{1}{4}\sum_{in=1}^{12} C(in)}{4}}{1 - 1/2}. \qquad (16)$$

Cases 1,2, and 3 correspond to three different ways of choosing the four nearest neighbors out of 12 (in one plane out of three orthogonal planes (100), (010), and (001)).

In our simulations, the sites with $C_A^{mean} = \frac{12}{16}$ correspond to the local LRO parameter close to 1 for the L1$_2$ structure, and those with $C_A^{mean} = \frac{8}{16}$ correspond to a local LRO parameter close to 1 for the L1$_0$ structure.

### 5.1. Main results, 0 < C < 0.25, Monte Carlo:

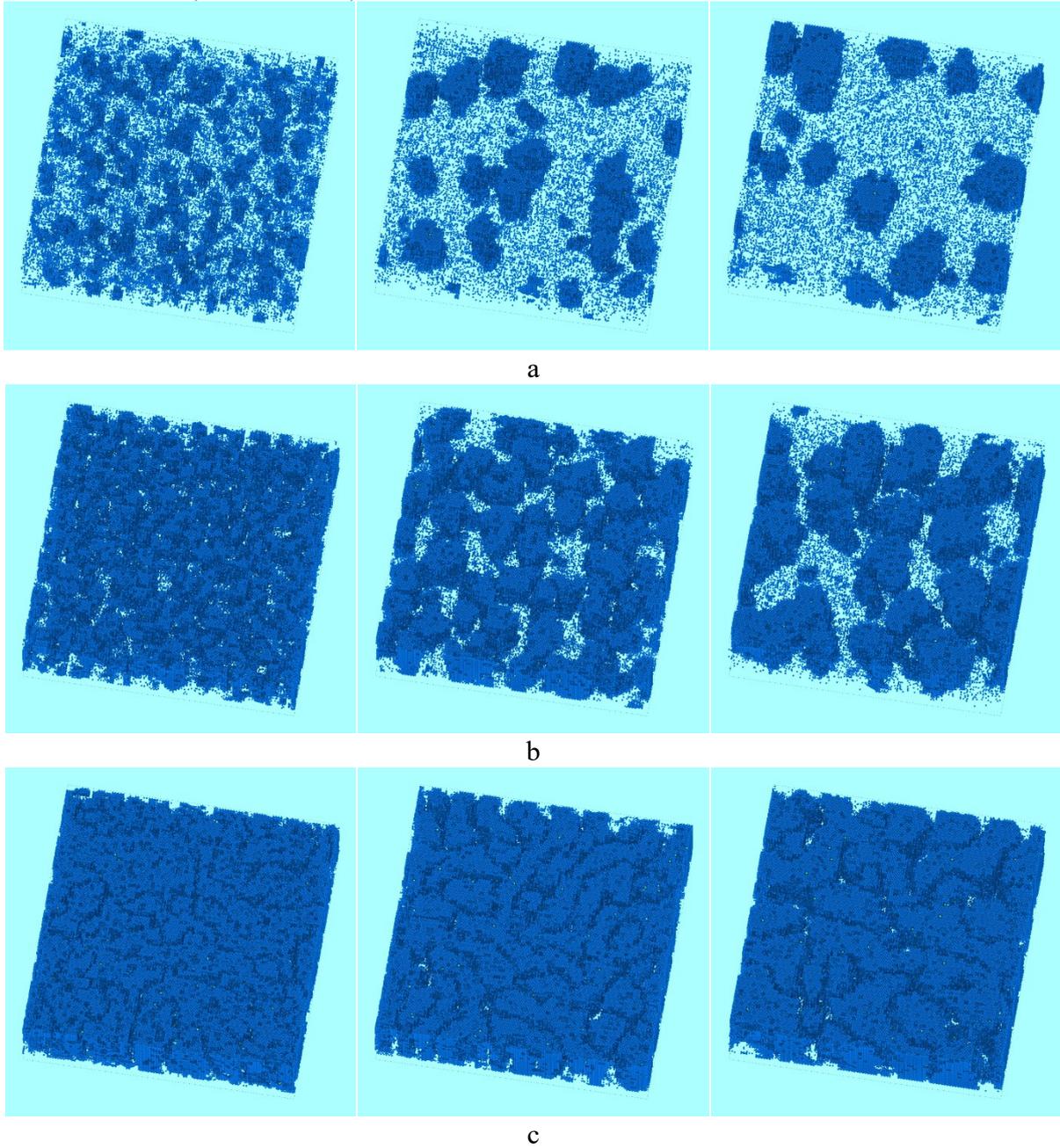

*Fig. 5. Patterns simulated by Monte Carlo for the case $0 < C^{dep} < \frac{4}{16}$ of decomposition into A+A3B phase (blue color, $C^{mean} = \frac{4}{16}$). Size: 200x200x55; Time: left - 100 MCSteps/plane; center - 1000 MCSteps/panel; right - 5000 MCSteps/plane. $C^{dep} = \frac{1}{16}$ (a), $\frac{2}{16}$ (b), $\frac{3}{16}$ (c)*

## 5.2. Main results, 0.25 < C < 0.50, Monte Carlo:

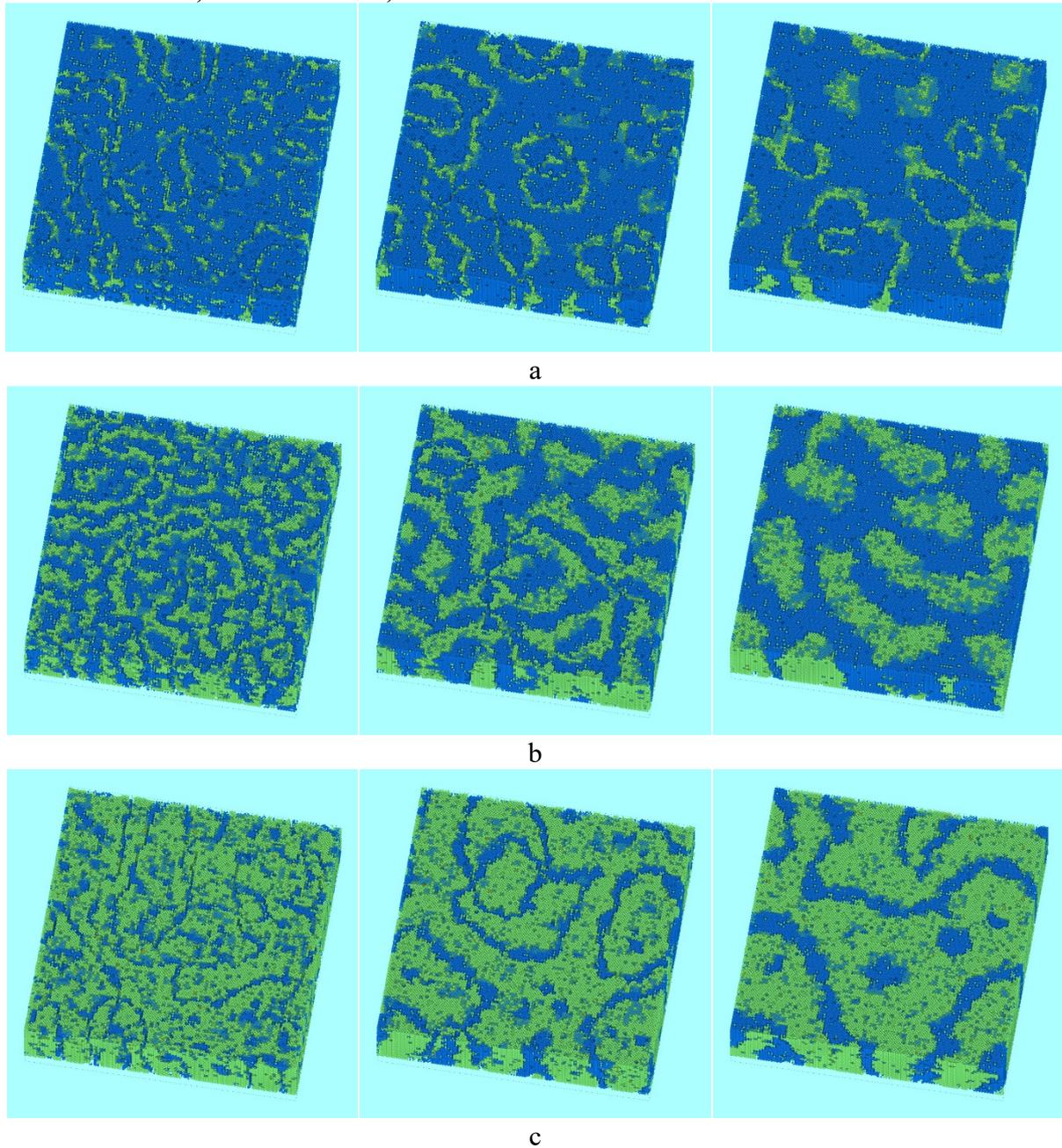

Fig. 6. Patterns simulated by Monte Carlo for the case $\frac{4}{16} < C^{dep} < \frac{8}{16}$ of decomposition into A3B-phase (blue color, $C_B^{mean} = \frac{4}{16}$ )+AB-phase (green color, $C_B^{mean} = \frac{8}{16}$ ). Size: 200x200x55; Time: left - 100 MCSteps/plane; center - 1000 MCSteps/panel; right - 5000 MCSteps/plane. $C_B^{dep} = \frac{5}{16}$ (a), $\frac{6}{16}$ (b), $\frac{7}{16}$ (c)

# 6. Conclusions.

1. Two possible cases of pattern formation due to phase separation during the co-deposition of the binary alloy from the vapor phase should be distinguished during simulation:
   A. Binary alloy with positive mixing energy demonstrating spinodal decomposition at sufficiently low temperatures, with compositions within the spinodal region of the surface layer.
   B. Binary alloy with negative mixing energy within the first coordination shell and (preferably) positive mixing energy within the second coordination shell, demonstrating strong long-range ordering within a narrow composition CB ranges around 1/4 (A3B-structure L12), 1/2 (AB-structure L10), and 3/4 (AB3-structure L12), and decomposition results in the formation of at least one ordered compound (depending on the composition of the deposition flux): a weak solution of B in A + A3B, a weak solution of A in B + AB3, A3B + AB, and AB + AB3.
2. The second case contains two subcases:
   B1. Decomposition into a weak solution and one ordered L12 compound (A + A3B) or (AB3 + B): $0 < C_{dep} < 0.25$ or $0.75 < C_{dep} < 1$;
   B2. Decomposition into an ordered L12 compound and another ordered compound L10 (A3B + AB) or (AB + AB3): $0 < C_{dep} < 0.25$ or $0.75 < C_{dep} < 1$.
3. Patterns steady-state morphology in 2D-sections can be quantified with the "home-made" topological parameter $P_{minority}$ (initially introduced in recent Letter [17]) for minority or 50/50 phase, which is close to 2 for rods (isolated spots in 2D-section), closer to 1 for layered structures (lamellae, zigzags, labyrinth), and closer to 0 for the net structure for minority, which is at the same time the "cell-structure" for majority phase. We observed only some elements of the net structure and only for the co-deposition with decomposition into two ordered compounds, A3B + AB.
4. One may also use the topological parameter for the majority. In limiting cases for three ideal structures (spots, lamellae, and net/cell), it was shown that $P_{majority}=1-P_{minority}$. In general, this seems not to be the case. Pattern formation in cases A and B1 seems to be similar: rod-like morphology at minority phase fractions less than approximately 35%, layered type for 45–50%, and some mixed patterns in between.
5. Let us go into some details about the new "net-like structure": Patterns formed via decomposition into two ordered compounds reveal one more tendency: that of a "net-like" structure for the minority phase, which simultaneously means a "cell-like" structure for the majority phase. The net-like morphology in the ideal case should provide a zero topological parameter P. At least for symmetric composition (6/16 or 14/16), when minority and majority fractions coincide, these curved nets and cells remind some flowers (for example, Orchidea).
6. The results of the simulation by the atomistic mean-field and Monte Carlo qualitatively coincide.

**Acknowledgements.**
The authors are grateful to PhD student Anastasiia Titova (ChNU), Prof. Janusz Sadowski (WU and Ensemble3, Warsaw), Prof. Oleksandr Kryshtal (AGH University, Cracow), and to Prof. Helen Zapolsky (Rouen University) for fruitful discussions of possible morphologies during the co-deposition of binary thin films. This work was supported by the ENSEMBLE3 Project (No. MAB/2020/14), which was carried out within the International Research Agendas Program (IRAP) of the Foundation for Polish Science, cofinanced by the European Union under the European Regional Development Fund, and the Teaming Horizon 2020 program (GA #857543) of the European Commission.